\documentstyle[preprint,aps,prl,graphicx,amssymb]{revtex}
\begin{document}

\title{Experimental evidence for high frequency transverse-like excitations in glasses}
\author{T.~Scopigno$^1$,
        E.~Pontecorvo$^1$,
        R.~Di~Leonardo$^1$,
        M.~Krisch$^2$,
        G.~Monaco$^2$,
        G.~Ruocco$^1$,
        B.~Ruzicka$^1$,
        F.~Sette$^2$
       }
\address{
         $^1$
         INFM and Dipartimento di Fisica, Universit\'a di Roma La
         Sapienza, 00185 Roma, Italy \\
         $^2$
         European Synchrotron Radiation Facility, BP 220, 38043
         Grenoble, France \\
        }

\date{\today}

\maketitle

\begin{abstract}
The dynamic structure factor of glassy and liquid glycerol has been measured
by inelastic X-ray scattering in the exchanged momentum ($Q$) region
$Q$=2$\div$23 nm$^{-1}$ and in the temperature range 80$\div$570 K. Beside the
propagating longitudinal excitations modes, at low temperature the spectra
show a second non $Q$-dispersing peak at $\hbar \Omega_T$$\approx$8.5 meV. We
assign this peak to the transverse dynamics that, in topologically disordered
systems, acquires a longitudinal symmetry component. This assignment is
substantiated by the observation that, in the liquid, this peak vanishes when
the structural relaxation time $\tau_\alpha$ approaches $\Omega_T^{-1}$, a
behavior consistent with the condition $\tau_\alpha \Omega_T^{-1}$$>>$1
required for the existence of a transverse-like dynamics in the liquid state.
\end{abstract}

\pacs{PACS numbers: 67.40.Fd, 62.60.+v, 67.55.Jd, 63.50.+x}



Understanding of the high frequency dynamics in the
glassy state has recently progressed thanks to the development of the
Inelastic X-rays Scattering (IXS) technique, allowing to measure
the dynamics structure factor, $S(Q,\omega)$, in the mesoscopic
momentum transfer, $Q$, range. This advance
established the presence of longitudinal acoustic-like excitations
\cite{phon} in topologically disordered systems. These excitations
exist up to momentum transfer values close to the first maximum of
the static structure factor, $S(Q)$, and have a characteristic
$Q$-dependent broadening related to the topological disorder
\cite{nondinor}. These results confirmed earlier predictions of
many Molecular Dynamics (MD) and Normal Mode Analysis (NMA)
simulation studies of disordered systems \cite{silica}. They are
also consistent with a recent extension of the Mode Coupling
Theory (MCT) to the glassy phase \cite{goetze}.

Despite these successes on the study of sound propagation at \textit{THz}
frequencies in topologically disordered systems, a satisfactory understanding
of the high frequency {\it transverse} dynamics in glasses is still missing.
Interestingly, beside the peaks associated with sound modes, theoretical and
numerical works also predict a second excitation in the $S(Q,\omega)$ or in
the related longitudinal current spectra \cite{goetze,waterexp,watersim}. This
feature has been suggested to indicate the existence of a transverse-like
dynamics in disordered systems, and it has been seen to survive also in the
liquid state  \cite{watersim}. The main reason supporting the transverse-like
nature of this second peak lies in the fact that it appears, at high $Q$'s, in
both the transverse and the longitudinal spectra, though being much more
intense in the former ones. Other general characteristics of this second
excitation are: \textit{i)} It appears in the spectra at $Q$ values larger
than approximately one half the value of the first maximum of the $S(Q)$;
\textit{ii)} It shows a weakly $Q$-dispersing behavior; and \textit{iii)} It
has an "harmonic" origin, as pointed out by its presence in both MD and NMA
simulations. In spite of these convincing numerical results, the firm
experimental observation of this feature in glasses is still missing. It would
be, therefore, of great interest to understand the behavior of this transverse
dynamics in the glassy state, at the glass transition, and eventually, the
conditions for its existence in the liquid state.

This Letter is dedicated to the experimental identification of this second
excitation in the high frequency dynamics of glycerol, an easily
experimentally accessible prototype glass forming system. Using IXS, we report
that in the glass, beside the sound modes, in the $S(Q,\omega)$ there is
indeed a second excitation at $\hbar\Omega_T$=$E_T$$\simeq$8.5 meV, showing up
at exchanged momenta larger than 7 nm$^{-1}$ with almost no $Q$ dispersion.
With increasing temperature this feature does not change energy, and
progressively disappears in the tails of the central peak, becoming no longer
observable above $\approx$500 K. At these high temperatures the width of the
central peak of the $S(Q,\omega)$, consistently with the extrapolation of the
temperature dependence of the viscosity, indicates that the structural
relaxation time, $\tau_\alpha$, approaches the inverse frequency of the second
peak. This observation suggests that this secondary peak (SP) does reflect the
transverse dynamics. Indeed, the transverse dynamics is fully relaxed in the
liquid at high temperatures, and appears as a vibrational contribution at low
$T$, when the condition $\Omega_T \tau_{\alpha}(T)$$>>$1 is fulfilled
\cite{BY}.

The experiment has been carried out at the very high resolution beam-lines
ID16 and ID28 of the European Synchrotron Radiation Facility. The instruments
consist of a backscattering monochromator and five independent analyzer
systems, held one next to each other with a constant angular offset on a 7 m
long analyzer arm. The utilized Si(11,11,11) configuration, with incident
photon energy of 21748 keV,  gives an instrumental energy resolution of 1.6
meV (FWHM) and an offset of 3 nm$^{-1}$ between two neighbor analyzers. The
momentum transfer is selected by rotating the analyzer arm. The energy scans
at constant $Q$ were performed by varying the monochromator temperature with
respect to that of the analyzer crystals. For the low temperature measurements
the glycerol sample, high purity anhydrous $C_3O_3H_8$, was put in a 20 mm
long pyrex cylindrical cell closed by 1 mm thick diamond windows with a 4 mm
opening. The cell was loaded inside an argon glove box and then placed in a
closed cycle cryostat. The high temperatures measurements have been performed
loading a 20 mm long nickel cylinder closed by two sapphire windows, (1 mm
thick, 8 mm diameter) and heated in a standard oven operated in vacuum.

\begin{figure}[p]
\centering
\includegraphics[width=.75\textwidth]{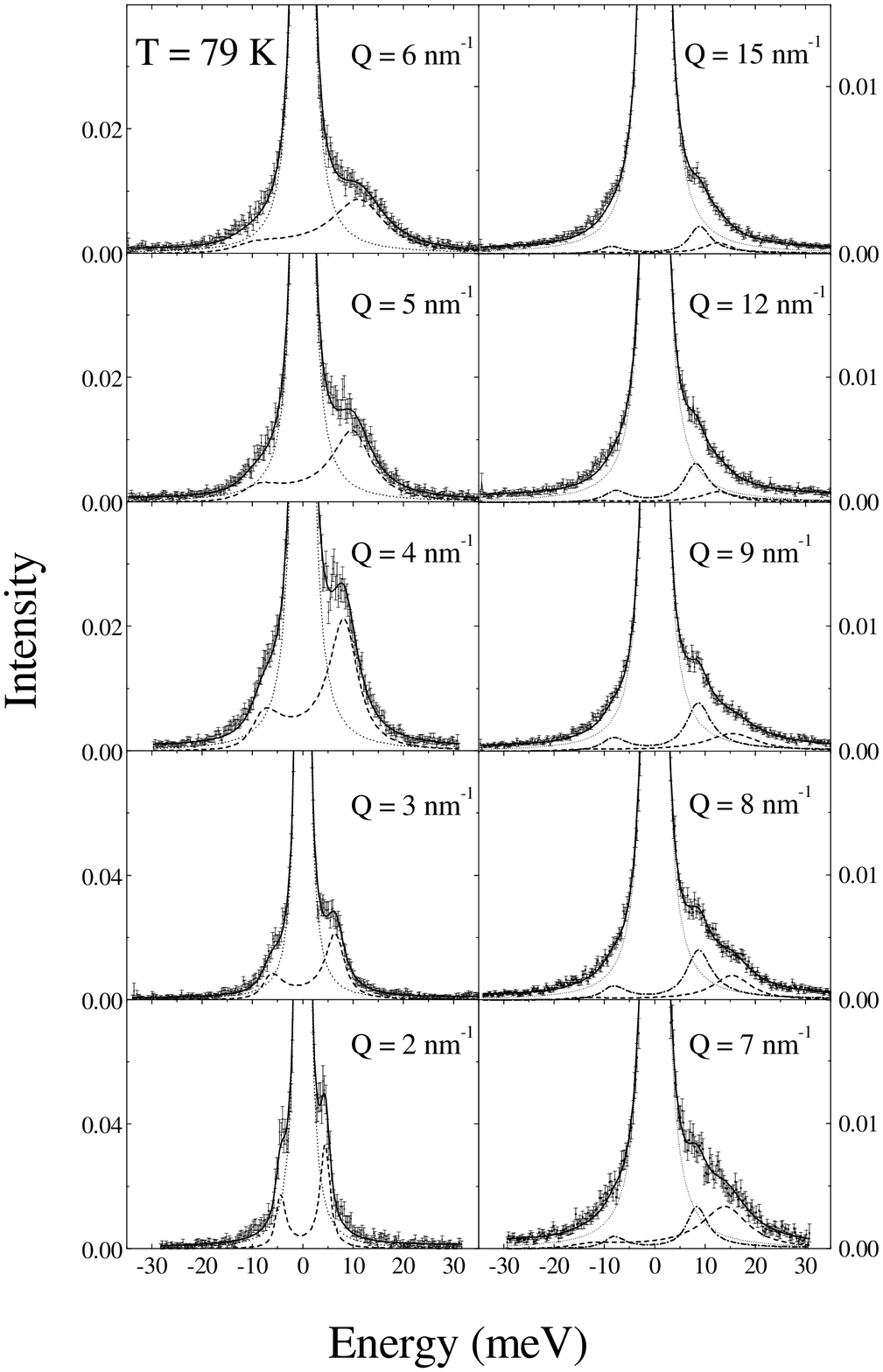}
\caption{IXS spectra of glassy glycerol at $T=79$ K for selected values of the
exchanged wavevector (dots with error bars). The total fitting result (full
line) is also reported, along with the genuine inelastic signal relative to the
longitudinal mode (dashed line). Above $Q \gtrsim 6$ nm$^{-1}$ (right column)
we included a SP with lorenzian shape in the fitting function (dotted-dashed
line). The instrumental resolution is also shown (dotted line).} \label{qpanel}
\end{figure}

The IXS data collected on the glycerol glass sample at $T$=79 K are reported
in Fig.~\ref{qpanel}. Beside an intense central line, the spectra show a $Q$
dependent inelastic signal. In the low $Q$ region, this is characterized by a
single feature whose energy position and width increase with $Q$. This
excitation is due to the propagating longitudinal acoustic modes, as already
established in previous works \cite{mascio,nondinor}. Thanks to the increased
statistical accuracy with respect to previous experiments, in the present
data, one observes also that at $Q$ larger than 7 nm$^{-1}$ a new weakly
dispersing feature appears at $\approx$10 meV. This feature, the secondary
peak shown in the right panel of Fig.~\ref{qpanel}, is particularly evident in
the spectra at the highest reported $Q$ values. The data have been fitted by
the convolution of the experimental resolution function with a model function,
weighted by the detailed balance, composed by: 1) a delta-function to account
for the central line, 2) a Damped Harmonic Oscillator (DHO) for the
longitudinal mode, and 3) a lorenzian for the SP, which is statistically
relevant only in the spectra at $Q \gtrsim 6$ nm$^{-1}$. This model well
represents the experimental data, as shown in Fig.~\ref{qpanel}.

\begin{figure}[p]
\centering
\includegraphics[width=.85\textwidth]{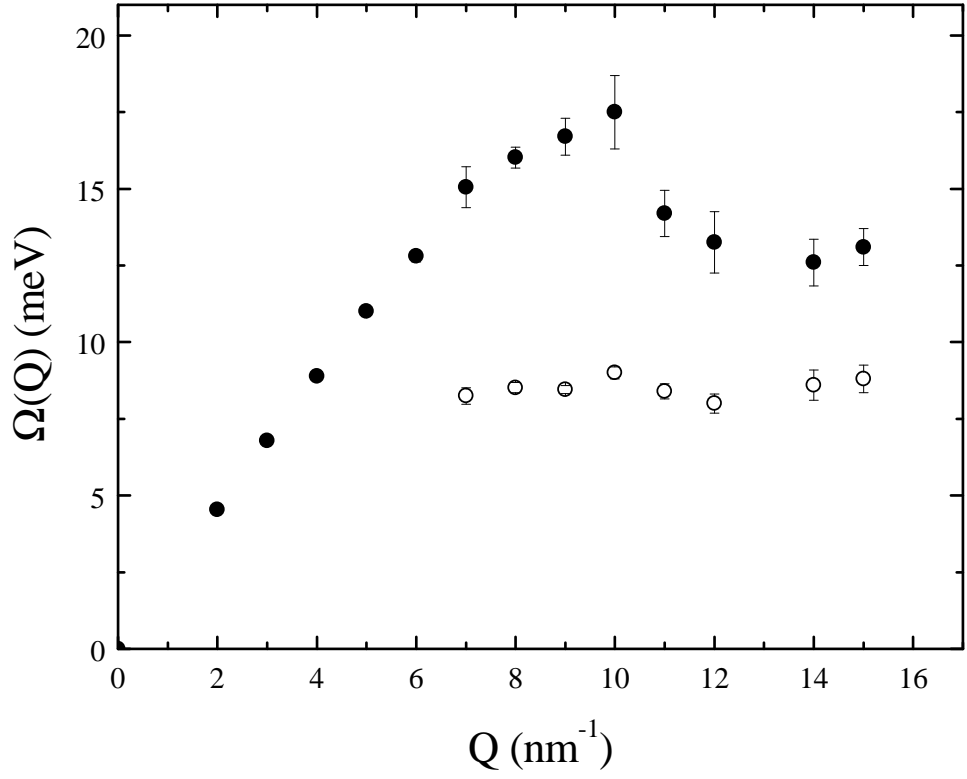}
\vspace{-8.2cm} \caption{Energy position of the longitudinal (full dots) and
transverse (open dots) acoustic modes at $T=79$ K, as determined by the
fitting procedure described in the text.} \label{disp}
\end{figure}

\begin{figure}[p]
\centering
\includegraphics[width=.75\textwidth]{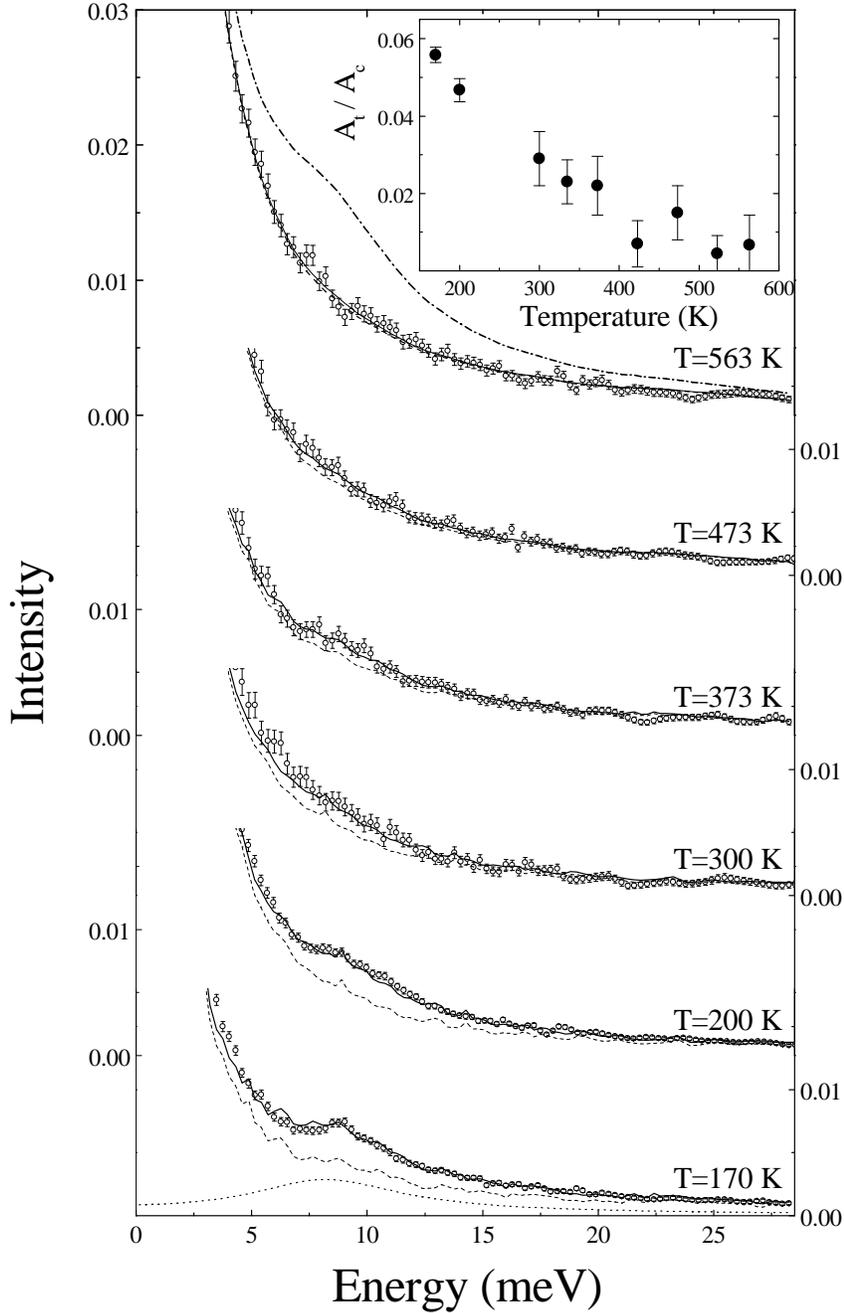}
\vspace{.5cm} \caption{Stokes side of the IXS spectra of glycerol at $Q=17$
nm$^{-1}$ at selected temperatures (open dots with error bars). Also reported
are the best fits to the data (full line) and the central peak contribution
(dashed line). For $T=170$ K we also report the inelastic signal (at the other
temperatures it simply scales for an amplitude factor), while at $T=563$ K we
show how the spectrum should appear if the SP were related to some
intramolecular mode. In the inset we report the inelastic (SP mode) to elastic
intensity ratio as determined by the fit.} \label{Tpanel1}
\end{figure}

In Fig.~\ref{disp}, we report the energy positions of the two features: the
longitudinal acoustic mode ($\Omega_L$) and the SP ($\Omega_T$). The slope of
$\Omega_L$ consistently reproduces, in the low-$Q$ linear dispersion region,
the correct value of the velocity of sound ($v_L$=3370$\pm$30 m/s). In the
high $Q$ region, we still observe a dispersion of $\Omega_L$ that, as
expected, has a maximum at a $Q$-value approximately half way from the maximum
of the $S(Q)$, and a minimum in the $Q$ region where the $S(Q)$ has a maximum.
This dispersion of $\Omega_L$ is consistent with the concept of pseudo
Brillouin zone in disordered systems, as discussed elsewhere \cite{umk}. The
SP is observed at the essentially $Q$-independent energy $\hbar
\Omega_T$=8.5$\pm$0.5 meV. This feature, as can be directly evinced from a
more detailed analysis of the data in Fig.~\ref{qpanel}, shows an interesting
$Q$-dependence in its intensity, which increases with increasing $Q$. This may
be due to various effects, among which the expected $Q^2$ dependence of the
cross-section of a non dispersing mode (localized or internal molecular mode),
and/or the possible $Q$-dependence of the longitudinal vs transverse symmetry
components of the eigenvectors associated to this mode. It is obviously
crucial to investigate these points to understand the origin of the SP. To
this purpose we studied this feature as a function of temperature in order to
investigate its evolution across the liquid-to glass transition ($T_g \approx
180$ K) and up to the normal liquid phase. This temperature dependence has
been studied at $Q$=17 nm$^{-1}$ from T=170 K (in the glassy phase) to T=570
K. The Stokes side of the spectra at selected temperatures is reported in
Fig.~\ref{Tpanel1}.

\begin{figure}[p]
\centering \vspace{-.2cm}
\includegraphics[width=.75\textwidth]{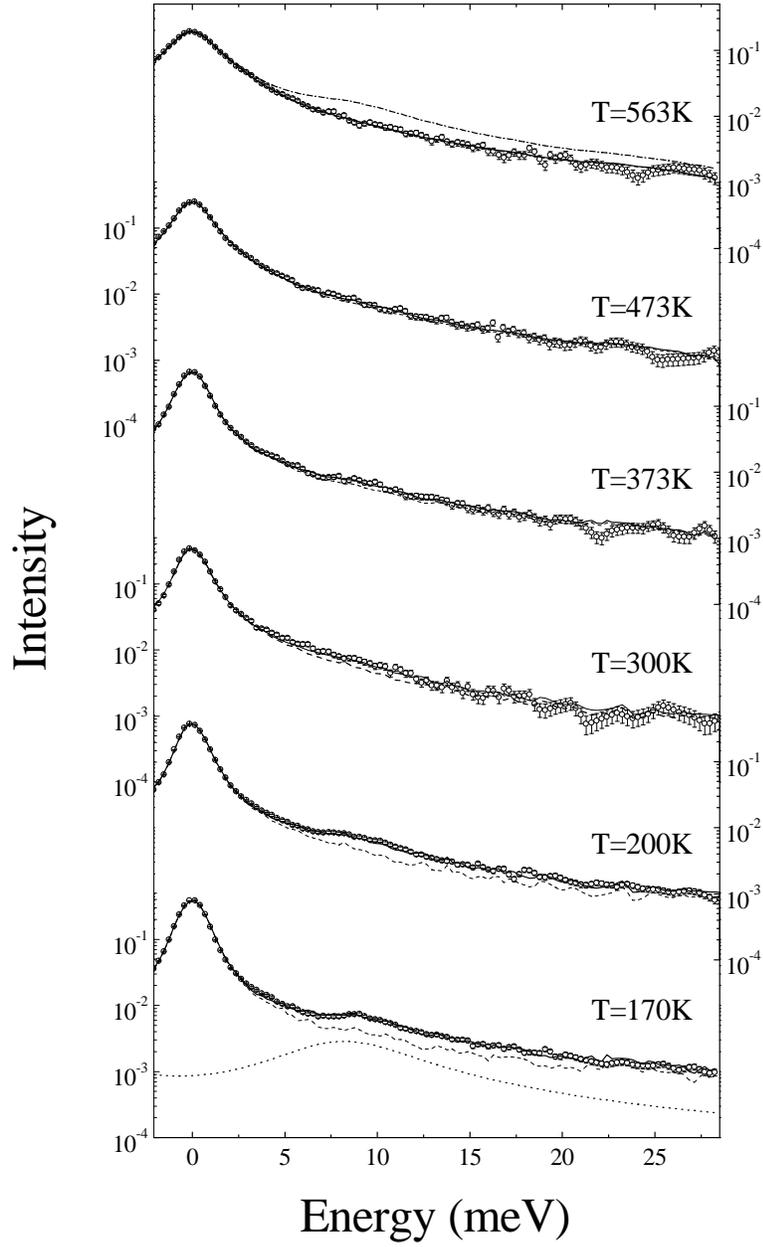}
\caption{Same as Fig. \ref{Tpanel1} but in logarithmic scale, to better
emphasize the broadening of the central peak above $T_g$.} \label{Tpanel2}
\end{figure}

The $Q$ value of 17 nm$^{-1}$ has been chosen in order to neglect the
contribution of the longitudinal mode, while having still the SP well
pronounced in the glass. The fitting function consists of a delta function for
the central line (replaced by a Cole-Davidson function for $T\gtrsim 400$ K,
when the intrinsic broadening can be distinguished by the mere resolution
linewidth), and a lorenzian function for the SP. Both spectral features show a
striking $T$-dependence: increasing $T$ the central line gets increasingly
broader and the SP becomes hardly visible already at $\approx$350 K. As long
as the SP is statistically significant one finds that its energy position and
width are $T$-independent, while its intensity decreases. The fits to the data
provide a quantitative description of these behaviors, and, as reported in the
inset of Fig.~\ref{Tpanel1}, the ratio between the integrated intensities of
the SP to the central peak sharply decreases at high temperatures. As shown in
Fig.~\ref{Tpanel2}, this behavior is opposite to the increase that one would
normally expect. Indeed: {\it i)} The central peak intensity should slightly
decrease, following the Debye-Waller $T$ dependence, and {\it ii)} The SP
intensity should increase following the Bose-Einstein (BE) statistics.

\noindent To emphasize further the statistical significance of the
SP intensity behavior, and to show that this feature is
not simply lost in the tails of the increasingly broader
central peak, we report at $T$=563 K the simulation
(dot-dashed line) of how the spectrum should look like under the
following assumptions: {\it i)} A central line with a constant
integrated intensity (the same as the spectrum at $T=170$ K) and
the observed width, {\it ii)} A SP obtained from the glass
spectrum at $T$=170 K, keeping constant the energy position and
width, and increasing the integrated intensity according to the BE
statistics. This simulation underlines the observed disappearance
of the SP at high $T$ beyond any statistical uncertainty.
The temperature dependence of the SP allows us then to exclude on a
general ground that this excitation is due to an internal molecular
mode. Indeed, if this were the case, one should expect a behavior
giving rise to the simulated spectrum in the upper panel of
Fig.~\ref{Tpanel1}.

\begin{figure}[p]
\centering
\includegraphics[width=.85\textwidth]{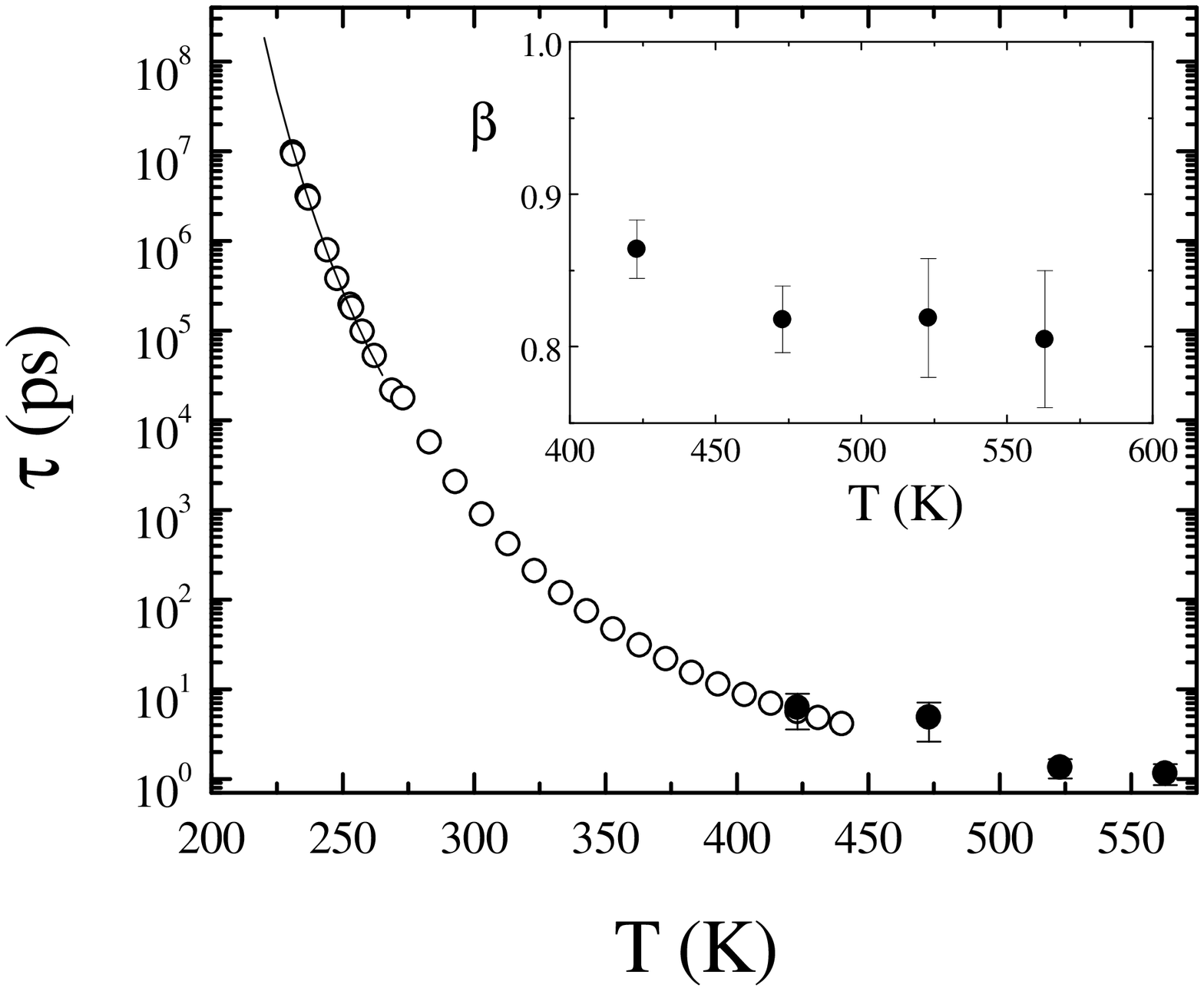}
\vspace{-7.8cm} \caption{Values of the structural relaxation time in glycerol
as function of temperature. Full dots: this work, determined in the liquid at
high temperature from the central peak broadening. Full line: light
scattering. Open dots: viscosity data scaled by an arbitrary factor. Inset:
values of the stretching exponent of the Cole-Davidson function used to
describe the central peak at high temperature} \label{visco}
\end{figure}

Finally, the broadening of the quasi-elastic line which appears in the
high-temperature spectra is quantified in Fig.~\ref{visco}, where we
report the structural relaxation time as obtained from the Cole-Davidson
contribution to the fitting function which we have used. These values
are in good agreement with light scattering and viscosity data
\cite{nelson,landolt}, testifying the reliability of the whole presented
data analysis.

On the basis of the reported results, we propose that the SP is due to the
transverse-like acoustic dynamics in a topologically disordered structure. In
fact, in the glycerol glass two excitation branches are present, as shown in
Fig.~\ref{disp}. This result is very similar to those obtained in MD
simulations, where indeed these two excitations have been assigned,
respectively, to the acoustic longitudinal-like and to the transverse-like
dynamics on the basis of the analysis of calculated longitudinal and
transverse current spectra. This assignment is also consistent with the
observation that the SP intensity increases with increasing $Q$. Indeed, on
decreasing the length scale, the local disorder forbids more and more the pure
polarization of the eigenvectors, thus leading to the mixing of the
longitudinal and transverse symmetries. Clearly, it is exactlty due to this
mixing that the longitudinal part of the transverse-like excitation comes out
in the $S(Q,\omega)$. This interpretation on the origin of the SP is
additionally supported by the behavior observed in the liquid phase. Indeed, a
vibrational transverse-like dynamics in the liquid state can only be supported
when the structural relaxation time is much longer than the transverse
vibration period (elastic limit), i.~e. when the local structure is frozen for
long enough time that a shear stress does not decay through relaxational
processes. Conversely, in liquids at high temperature, where the condition
$\Omega_T \tau_\alpha$$>>$1 is no longer fulfilled, the vibrational transverse
dynamics is no longer allowed and the SP disappears. It is important to point
out that this behavior is distinctly different from that of the longitudinal
acoustic excitations. In fact - as experimentally observed - these latter ones
keep their vibrational nature independently of the value of $\Omega_L
\tau_\alpha$, which only affects the value of the sound velocity and
absorption.

In conclusion, we report experimental evidences of the existence of a
transverse-like dynamics in glasses and liquids at THz frequencies. This
dynamics gives rise to a well defined and $non$ \textit{Q}-dispersing
excitation, consistently with MD simulations. We expect that this feature is a
general property of disordered matter, and that it should be universally
observed. Indications in this direction come not only from MD simulations
\cite{watersim,tesiR}, but also from experimental data on glassy SiO$_2$ and
liquid water \cite{JPC,waterexp}. Future work on the dynamics in disordered
systems should consider the contribution of this high frequency
transverse-like dynamics to the vibrational density of states. This could be
particularly relevant in the case of glasses, due to the possible contribution
and/or relation of this transverse-like dynamics to the well known low
temperature anomalies.

We acknowledge C.~Henriquet, H.~Mueller, and R.~Verbeni to help in the sample
handling.

\begin{references}
\bibitem{phon}
F. Sette, M. Krisch, C. Masciovecchio, G. Ruocco, and G. Monaco,
Science {\bf 280}, 1550 (1998); O. Pilla, et al., Phys. Rev. Lett.
{\bf 85}, 2136 (2000) and references therein.

\bibitem{nondinor}
G. Ruocco, et al., Phys. Rev. Lett. {\bf 83}, 5583 (1999).

\bibitem{silica} S.N. Taraskin and S.R. Elliot, Europhys. Lett.
{\bf 39}, 37 (1997), ibid. Phys. Rev. B {\bf 56}, 8605 (1997).
M.C. Ribeiro, M. Wilson and P.A. Madden, J. Chem. Phys. {\bf 108},
9027 (1998), ibid. J. Chem. Phys. {\bf 109}, 9859 (1998). R.
Dell'Anna, G. Ruocco, M. Sampoli, and G. Viliani, Phys. Rev.
Lett.  {\bf 80}, 1236 (1998). J.L. Feldman, P.B. Allen and S.R.
Bickham, Phys. Rev. B {\bf 59}, 3551 (1999). J. Horbach, W. Kob
and K. Binder, Eur. Phys. J. B {\bf 19}, 531 (2001).

\bibitem{goetze}
 W. Goetze, and M. R. Mayr, Phys. Rev. E  {\bf 61}, 587 (2000).

\bibitem{waterexp}
F. Sette, et al., Phys. Rev. Lett. {\bf 77}, 83 (1996).

\bibitem{watersim}
M. Sampoli, G. Ruocco, and F. Sette, Phys. Rev. Lett. {\bf 79},
1678 (1997).

\bibitem{BY}
 J. P. Bonn, and S. Yip, {\it Molecular Hydrodynamics}, McGraw-Hill, New York,
1980.

\bibitem{mascio} C. Masciovecchio, et al., Phys. Rev. Lett. {\bf 76}, 3356 (1996)

\bibitem{umk} T.~Scopigno, M.~D'Astuto, M.~Krisch, G.~Ruocco, F.~Sette,
Phys. Rev. B {\bf 64}, 012301 (2001).

\bibitem{nelson} D.M. Paolucci, K.A. Nelson, J. Chem. Phys. {\bf
112}, 6725 (2000).

\bibitem{landolt} H. Landolt, R. Bornestein sec. 25-1, vol. II/5a,
VI edition.

\bibitem{tesiR} R. Dell'Anna, Ph.D. Thesis, L'Aquila (1997), unpublished.

\bibitem{JPC} G. Ruocco, F. Sette, J. Phys. C {\bf 13}, 9141 (2001).

\end{references}
\end{document}